\documentclass[aps]{revtex4}
\usepackage{graphicx}
\usepackage{epsf}
\usepackage{wrapfig}
\usepackage{epsfig}
\usepackage{epsfig}
\def\Vec#1{\mbox{\boldmath $#1$}}
\def\beq{\begin{equation}}
\def\eeq{\end{equation}}

\def\beqy{\begin{eqnarray}}
\def\eeqy{\end{eqnarray}}
\begin{document}
\title{Realistic Nuclear Wave functions and Heavy Ion Collisions}
\author{M. Alvioli}
\altaffiliation{Also at: Department of Physics, University of Perugia and
      INFN, Sezione di Perugia, Via A. Pascoli, I-06123, Italy}
\affiliation{104 Davey Lab, The Pennsylvania State University,
  University Park, PA 16803, USA}
\author{C. Ciofi degli Atti}
\affiliation{Department of Physics, University of Perugia and
      Istituto Nazionale di Fisica Nucleare, Sezione di Perugia,
      Via A. Pascoli, I-06123, Italy}
\author{M. Strikman}
\affiliation{104 Davey Lab, The Pennsylvania State University,
  University Park, PA 16803, USA}
\begin{abstract}
We discuss the implications of recent experimental evidence of nuclear Short
Range Correlations (SRCs) on the modeling of the wave function of complex
nuclei. We perform a calculation of potential energy contributions of $pp$
and $pn$ pairs in nuclei, showing that the presence of strong tensor
correlations produce a ratio between the twos of about 1/9, which strongly
deviates from the combinatorial counting of the number of pairs. We also discuss
implications for the production of Monte Carlo configuration for the simulation
of nuclear reactions involving complex nuclei.
\end{abstract}
\maketitle

In the past decade, various measurements have been unambigously identified
two-nucleon Short Range Correlations, studied their structure and related
them to the underlying basic short range Nucleon-Nucleon (NN) interaction
\cite{Egiyan:2006}-\cite{Piasetzky:2006ai}. In \cite{Egiyan:2006} two- and
three-body correlations were studied in large nuclei and related to deuteron
and $^3He$ measurements; the $(p,2p+n)$ \cite{Tang:2002ww} experiment studied
the directional correlation between proton and neutron momenta, while in
\cite{Piasetzky:2006ai} it was shown that about $90\%$ of high momentum
protons are correlated with a neutron; in \cite{Subedi:2008zz} high momentum
protons in $(e,e^\prime p)$, $(e,e^\prime pp)$ and $(e,e^\prime pn)$ were
investigated, finding a recoling, back-to-back proton in $10\%$ of the events and
a neutron in $90\%$ of the events, consistently with \cite{Piasetzky:2006ai}.
After these experiment, several nuclear theory groups showed that the measured
ratio is a strong indication of the operation  of an NN tensor force in the pair
at the nucleon separations and relative momenta
studied \cite{Alvioli:2007zz}, \cite{Schiavilla:2006xx}. Recently, an effective
many-body approach for the description of ground-state wave functions have been
adopted for complex nuclei by the Perugia group, relying on a cluster expansion
for the calculation of expectation values of quantum operators on the ground state
wave functions with realistic forces \cite{Alvioli:2005cz}.
In this work, we investigated the effect of correlation on the fraction of potential
energy in the nucleus which is carried by the different $pn$ and $nn$ pairs; moreover,
we investigated the possibility of inclusion of full state-dependent correlations in
nuclear configurations for the simulation of NA and AA collision at high energies.

\begin{figure}[!ht]
\vskip -0.9cm
\centerline{\includegraphics[width=19cm]{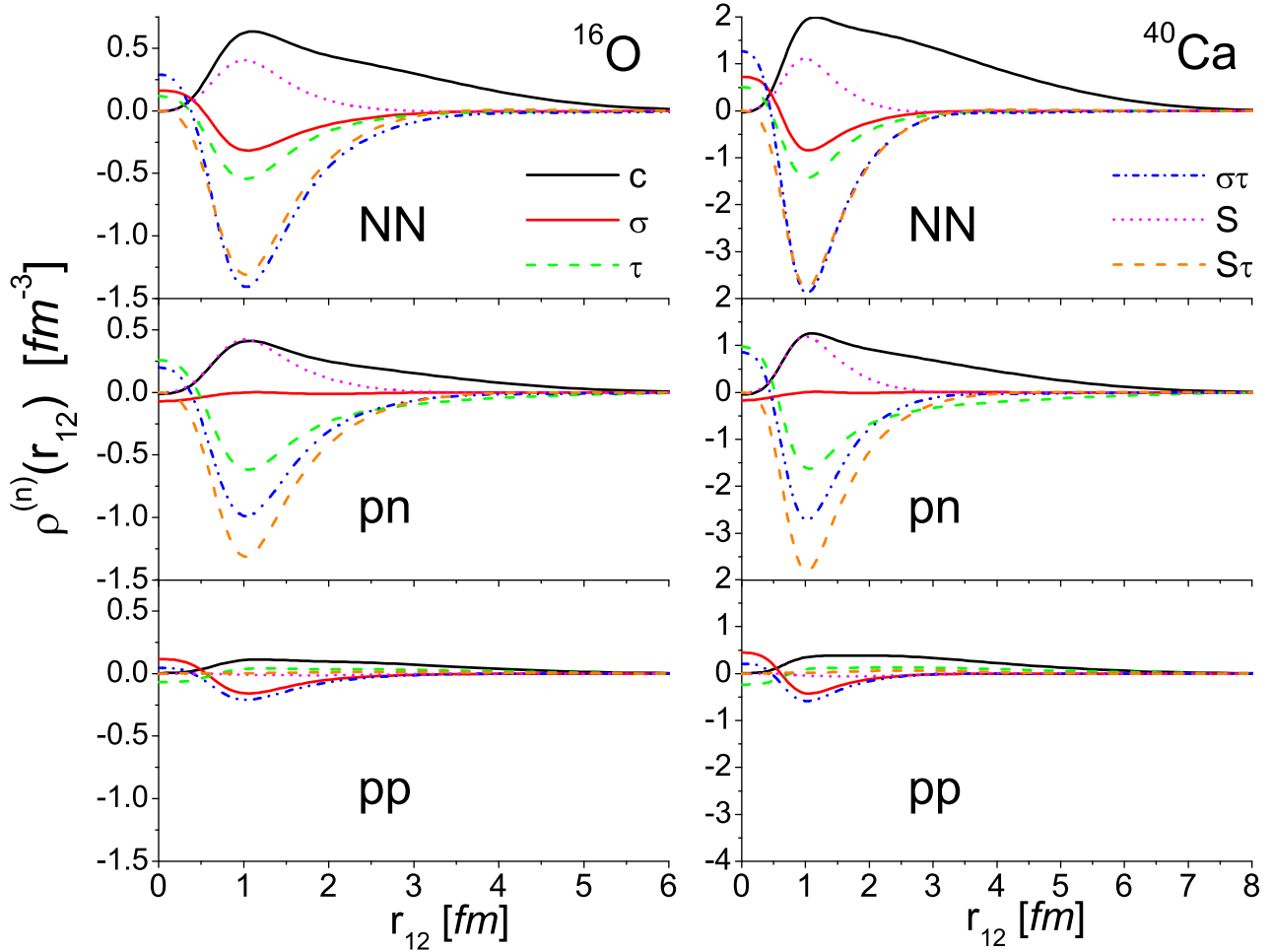}}
\vskip -1.0cm
\caption{The radial, state dependent two-body density matrices (Eq.~\ref{eq5})
  of $^{16}O$ (\textit{left column}) and $^{40}Ca$ (\textit{right column}),
  as defined in Eq.~(\ref{eq3});
  \textit{Top}: the total Nucleon-Nucleon density;
  \textit{Center}: the proton-proton density;
  \textit{Bottom}: the neutron-proton density.}
\label{Fig1}
\end{figure}

We can calculate the potential contribution to the ground state energy according to
\beq
\label{eq1}
\langle V\rangle\,=\,\sum^A_{i<j}\langle\,\hat{v}(r_{ij})\,\rangle\,=
\,\sum^A_{i<j}\langle\,\sum^6_{n=1}v^{(n)}(r_{ij})\,\hat{O}^{(n)}_{ij}
\,\rangle\,=\,\frac{A(A-1)}{2}\,\sum^6_{n=1}\,\int d\Vec{r}_1d\Vec{r}_2
\,\rho^{(2)}_n(\Vec{r}_1,\Vec{r}_2)\,v^{(n)}(r_{12})\,.
\eeq
Here $\rho^{(2)}_{n}(\Vec{r}_1,\Vec{r}_2)$ is the state-dependent two-body
density matrix
\beq
\label{eq3}
\rho^{(2)}_n(\Vec{r}_1,\Vec{r}_2)\,=\,\int \prod^A_{j=3}d\Vec{r}_j
\,\psi^\star(\Vec{r}_1,...,\Vec{r}_A)\,\hat{O}^{(n)}_{12}
\,\psi(\Vec{r}_1,...,\Vec{r}_A)\,,
\eeq
which we have evaluated within the cluster expansion method with
$\psi(\Vec{r}_1,...,\Vec{r}_A)=\prod^A_{i<j}\hat{f}(r_{ij})\phi(\Vec{r}_1,...,\Vec{r}_A)$,
$\phi$ being the independent particle model wave function,
and $\hat{O}^{(n)}_{12}$ is the operator
\beq
\label{eq4}
\hat{O}^{(n)}_{12}\,\in\,\left\{\hat{1},\Vec{\sigma}_1\cdot\Vec{\sigma}_2,
\hat{S}_{12}\right\}\,\otimes\,\left\{\hat{1},\Vec{\tau}_1\cdot\Vec{\tau}_2\right\}\,,
\eeq
acting between particles $1$ and $2$, present in both the nucleon-nucleon potential
and in the ground state wave function.
The two-body density appearing in Eq.~(\ref{eq1}) can be splitted within our approach
into the contributions due to proton-proton, proton-neutron and neutron-neutron pairs
and the potential energy can thus be splitted into the corresponding contributions.
The radial two-body density
\beq
\label{eq5}
\rho^{(2)}_n(r_{12})\,=\int d\Vec{R}\,\rho^{(2)}_n
\left(\Vec{r}_1=\Vec{R}+\frac{1}{2}\Vec{r}_{12}\,,
\,\Vec{r}_2=\Vec{R}-\frac{1}{2}\Vec{r}_{12}\right)
\eeq
is shown in Fig. \ref{Fig1} for $^{16}O$ and $^{40}Ca$, for
$n=c,\sigma,\tau,\sigma\tau,S,S\tau$ in Eqs. (\ref{eq3}), (\ref{eq4}) and (\ref{eq5}).
We have calculated the $pp$ and $pn$ contributions to potential energy for $^{16}O$
and $^{40}Ca$, using the density of Eq.~(\ref{eq3}) and the Argonne $AV8^\prime$
potential, obtaining for the following results. The contribution from
\textit{central} correlations is small and in this case the probabilities
from $pp$ and $pn$ pairs are exactly proportional to the fractions
estimated combinatorically, namely $Z(Z-1) / A(A-1)$ and $2ZN / A(A-1)$, respectively,
giving $P(pp)=23\%$ and $P(pn)=53\%$ for $^{16}O$.
In the case of full correlation, this proportionality does not hold anymore, and
we find $P(pp)=8\%$ of the total and $P(pn)=83\%$ of the total, or $P(I=0)=74\%$
and $P(I=1)=26\%$, $I$ being the total isospin of the pair. The corresponding
calculations for $^{40}Ca$ give practically the same results.

\begin{figure}[!htp]
\vskip 0.5cm
\leftline{\hspace{0cm}
  \epsfxsize=10.0cm\epsfbox{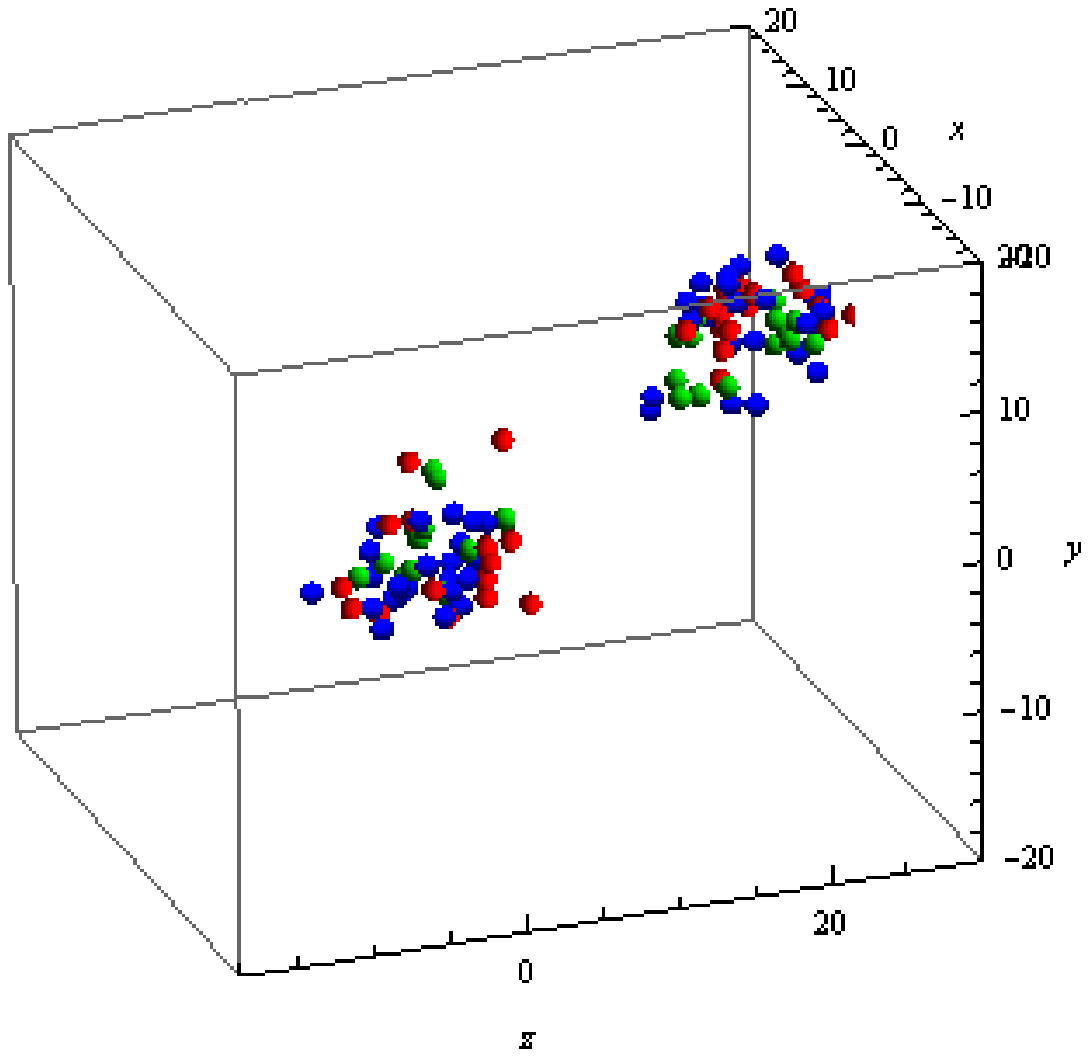}}
\vskip -9.0cm
\rightline{
  \epsfxsize=8.0cm\epsfbox{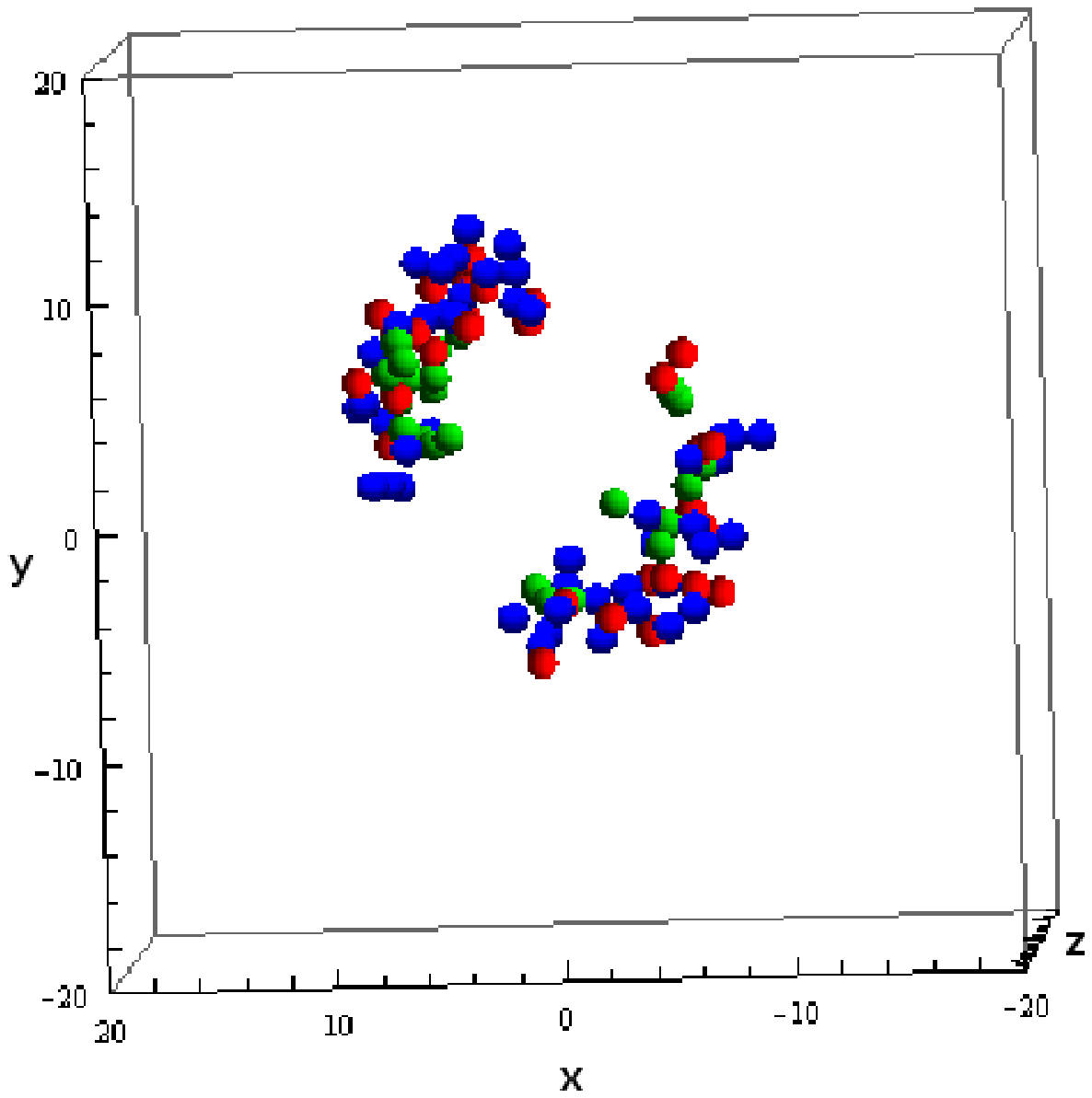}\hspace{0cm}}
\vskip 1.0cm
\caption{The spectator systems after a $Pb-Pb$ collision along the z direction.
  \textit{Left}:
  side view; \textit{right}: view from behind. Red and blue spheres are
  protons and neutron, respectively, while the green ones are those nucleons
  which belonged to a correlated pair, before the interaction, their correlated
  partner being among the interacting nucleons (hidden, in this figure). The
  axes units are in $fm$ and the dimension of the spheres are taken as the
  rms charge radius of the proton; Glauber parameters correspond to RHIC energies.
  Animations are available \cite{download} along with
  the configurations used for the colliding nuclei.
}\label{Fig2}
\end{figure}

When simulating the collision of two heavy ions, one starts by constructing some
picture of the two involved nuclei, with the positions of the nucleons are sampled
according to a density distribution function; this approach completely ignores the
structure of the wave function of the nucleus, which is a highly complicated object
depending on the positions, momenta, spin and isospin states of A nucleons.
In Ref.~\cite{Alvioli:2009ab} it was developed a Monte Carlo code to provide
a more realistic implementation of the nuclear wave function, including
\textit{central} correlations, and providing a good description of two-body
densities of the nucleus as compared with the ones obtained within an independent
particle model. From this work, we know that basic quantities as the potential
energy of the nucleus are strongly affected by realistic SRCs, therefore it appears
mandatory  to include \textit{state-dependent} correlations in the approach of
Ref.~\cite{Alvioli:2009ab}.
To this end, the first improvement to the mentioned approach was to distinguish
protons and nucleons, and to implement state-dependent correlations between
first-neighbor nucleons. With the newly developed configurations \cite{download}
we will be able to investigate various aspects of AA collision \cite{Alvioli:NEW};
Fig. \ref{Fig2} shows the spectator systems after a $Pb-Pb$ collision, the
NN scattering being treated within the Glauber theory. Correlated nucleons,
which will be emitted with high momentum, are depicted in a different color.

\textit{M.A. is supported by DOE grant under contract DE-FG02-93ER40771.
We thank the HPC-Europa2 Consortium (project number: 228398), with
the support of the EC - Research Infrastructure Action of the FP7 -
and EPCC for the use of computing facilities.}


\end{document}